\documentclass[11pt]{article}
\usepackage{makeidx}
\usepackage{multirow}
\usepackage{multicol}
\usepackage[dvipsnames,svgnames,table]{xcolor}
\usepackage{graphicx}
\usepackage{epstopdf}
\usepackage{ulem}
\usepackage{hyperref}
\usepackage{amsmath}
\usepackage{amssymb}
\author{Microsoft Office User}
\title{}
\usepackage[paperwidth=595pt,paperheight=841pt,top=113pt,right=70pt,bottom=76pt,left=70pt]{geometry}

\makeatletter
	{\par\setlength{\parindent}{#3}
	\setlength{\leftmargin}{#1}       \setlength{\rightmargin}{#2}%
	\advance\linewidth -\leftmargin       \advance\linewidth -\rightmargin%
	\advance\@totalleftmargin\leftmargin  \@setpar{{\@@par}}%
	\parshape 1\@totalleftmargin \linewidth\ignorespaces}{\par}%
\makeatother 

\begin{document}

{\raggedright
{\small \textit{Journal of Physics Conference Series}~\textbf{1251}:012028 (June
2019)}
}

{\raggedright
{\small  DOI:~ 10.1088/1742-6596/1251/1/012028}
}

\begin{center}
\textbf{{\LARGE Magnetic monopole as the shadow side of the electric
charge}} \footnote{\textit {Proceedings of 11th International Symposium Honoring
Noted Mathematical Physicist Jean-Pierre Vigier.} 

ADVANCES IN FUNDAMENTAL PHYSICS, Liege, Belgium, 6 -- 9 August 2018. 

Editors: R. l. Amoroso,  D. Dubois and P. Rowlands.}  

\end{center}

\begin{center}
\textbf{Volodymyr Krasnoholovets}
\end{center}

\begin{center}
Institute of Physics, Natl. Acad. Sci. of Ukraine, 46 Nauky St., Kyiv 03028,
Ukraine
\end{center}

\begin{center}
E-mail: krasnoh@iop.kiev.ua
\end{center}

\textbf{Abstract}.\textbf{ } It is shown that a magnetic monopole appears as the tension state of the primary electric charge at its motion through each section of the path equal to the particle's de Broglie wavelength. This conclusion is followed from a submicroscopic consideration of particles and their motion in the framework of the theory of physical space in the form of a tessellattice developed by Michel Bounias and the author. The periodical change of the particle's charged state to its monopole state can easily be introduced in the conventional Maxwell equations and the magnetic monopole automatically shows up in the structure of Maxwell's equations. The monopole is also presented in any quark system as a quark obeys dynamics that are also characterized by the appropriate de Broglie wavelength and hence the electric charge changes periodically to the magnetic monopole. A (anti)neutrino emerges as the typical electron's monopole. When the charged particle becomes the monopole, it also loses its mass (the mass passes to the particle's inerton cloud) and thus the neutrino is a massless particle, or more correctly massless monopole.

\section*{1. Introduction}

Here are Maxwell's equations (in the SI base units):
                      
\begin{eqnarray}
\label{eq1}
 {\boldsymbol \nabla} \cdot {\bf E}  &=& \frac{\rho}{\epsilon_{0}} ,   \\
 {\boldsymbol \nabla} \cdot {\bf B} &  = & 0    \\
 {\boldsymbol \nabla} \times {\bf E} &= &- \frac {\partial {\bf B}} {\partial t},    \\
{ \boldsymbol \nabla} \times {\bf B} & = & \epsilon_0 \mu_0 \frac{\partial {\bf E}} {\partial t} + {\mu_0} {\bf j}.
\end{eqnarray}

Eq. (1) reflects Gauss' law, Eq. (3) characterizes Faraday's law and Eq. (4) describes Ampere's law. The symbols in Eqs. (1) to (4) are: {\bf E} is the dynamic electric field (the dimensionality is [kg$\cdot$m$\cdot$s$^{-3}$$\cdot$A$^{-1}$]), {\bf B} is the magnetic induction (the dimensionality is [A$\cdot$m$^{-1}$]),  is the density of the electric charge (the dimensionality is [C$\cdot$m$^{-3}$]), j is the density of the electric current (the dimensionality is [A$\cdot$m$^{-2}$]), and $\epsilon_0 = 8.85418782 \times 10^{-12}$ m$^{-3}$$\cdot$kg$^{-1}$$\cdot$s$^4$$\cdot$A$^2$ and $\mu_0 = 1.25663706 \times 10^{-6}$ m$\cdot$kg$\cdot$s$^{-2}$$\cdot$A$^{-2}$  are electric and magnetic constants, respectively.

Although Maxwell?s equations (1)?(4) correctly describe electromagnetic phenomena, the absence of a magnetic charge in them haunts many researchers ? why are the equations not symmetric? It seems magnetic fields can be understood to be caused by magnetic charges. But where is the magnetic charge? 

Lochak [1] notes that the hypothesis of separated magnetic poles is very old and was considered already by Coulomb and Maxwell. Dirac [2] suggested the existence of isolated magnetic poles (monopoles). His monopole $g$ should be connected with the electric charge by the expression $hc/(eg)=2$, which means the further quantization of the fundamental electric charge $e$.

In present days the magnetic monopole is treated as an important fundamental particle (see, e.g. Refs. 1-4) whose existence is postulated based on the duality symmetry. Shnir [4] emphasizes: `` ... our imperfect (at least at low-energy scale) world is full of nasty broken symmetries. This has impelled physicists to try to understand how this happens."

By introducing the idea of magnetic monopoles, Maxwell's equations become symmetrical with respect to the magnetic and electric fields, but they still remain unsymmetrical as far as the scalar and the vector potentials are concerned. In some works [5, 6] the electric and magnetic fields are redefined in terms of some new scalar and vector potentials, respectively, as a result of which the Maxwell?s equations become symmetrical with respect to the potentials too. 
      
Theorists suggest (see, e.g. Ref. 4) an idea of a dyon structure of particles in which the monopole is moving in the field of the electric charge. Lochak's [1] own theory is based on the fact that the massless Dirac equation admits a second electromagnetic coupling, deduced from a pseudo-scalar gauge invariance; so the equation obtained has the symmetry laws of a massless leptonic, magnetic monopole, able to interact weakly. 
      
On the other side, Gonalo and Zich [6] argue that {\bf E} and {\bf B} fields exhibit very different natures, which especially appears in the N-D extension of cross product and this allows them to reinterpret magnetic field {\bf B} in such a way that denies magnetic monopoles at all.
     
Although so far no experiments have directly revealed monopole magnetic particles, some physicists still believe in their existence hoping that these particles can experimentally be detected in future. 

\section*{2. Magnetic monopoles}	

Of course it is impossible not to notice that Maxwell's equations (1)?(4) possess the apparent asymmetry of describing electrical and magnetic phenomena. Electricity and magnetism enter the equations unequally. The second equation directly indicates that there are no magnetic charges in nature. What is more, no stationary magnetic charges. Electric fields are created either by electric charges or by varying magnetic fields, while magnetic fields are created only by electric current and varying electric fields. 
      
 If the magnetic monopole g could exist, then in the SI base units its dimensionality is [m$^{-2}$$\cdot$A], though the dimensionality of the electric charge $e$ is [s$\cdot$A]. The dimensionality [A] of electric current {\it I} is related to [C$\cdot$s$^{-1}$]. Looking at the dimensionality of $g$, we can see that it exactly corresponds to the current density, $[g]=[j] = [\rho{\kern 1pt} v] \equiv $ m$^{-2}\cdot$A. Does it mean that the magnetic monopole vector is associated with the current density vector and hence magnetic monopoles are present in Maxwell's equations? That is,                                                                                             

\begin{equation}\label{5}
{\bf g} = {\bf j} = \rho {\bf v} ,                            
\end{equation}    
where \textbf{v } is the vector velocity of the electric charge.

The answer to the question above cannot be found in the framework of both orthodox quantum electrodynamics and quantum electrodynamics. The quantum theory only uses main postulates of classical electrodynamics and does not consider the nature of the charge and its behavior at a subatomic scale. Electromagnetic properties are associated with photons, carriers of the electromagnetic field but what is the photon? -- any quantum theory cannot answer this either.  

To answer these complex questions, we must plunge into the network of such a notion as the physical space in which all physical phenomena occur, such as the birth of particles, their movement, interaction and transformation. 

Theoretical attempts to construct monopoles in the frameworks of physical mathematics (Dirac, 't Hooft, Polyakov and others) failed as no such exotic monopoles have been found experimentally. Besides, such approaches are only similar to mathematical physics, but they are very far from the natural realities by definition.

\section*{3. Magnetic monopole in the tessellattice}

       On the other hand, the solution to the problem of the existence of magnetic monopoles can come from a detailed consideration of the notion of real physical space. Namely, we may start from a submicroscopic concept that discloses a mathematical construction of our ordinary space at a submicroscopic scale, i.e. Planck's scale. Such a theory was developed by Bounias and the author in a series of works (see e.g. monograph [7]). The theory shows that space represents a mathematical lattice of primary topological balls. The lattice was named the tessellattice by Michel Bounias [8] and the size of its cell (a topological ball) can be associated with the Planck?s size. 
       
The tessellattice allows a number of structures and properties. The appearance of a canonical particle is associated with a change in the configuration of a cell. If a cell is contracted following a fractal law, this means the appearance of a massive particle (a lepton like an electron). If a cell is inflated, this means the appearance of a quark. The electric and magnetic properties are related to the configuration of the surface of the cell. So, the physical notion of mass relates to a mathematical ratio of the initial volume of a degenerate cell to the volume of the fractally deformed cell.   
       
The created particle induces a deformation coat around itself, which is extended to the distance equal to the particle?s Compton wavelength $\lambda_{\rm Com}=h/(mc)$. The deformation coat screens the local deformation generated by the created particle from degenerate cells of the tessellattice, i.e. up to the distance $\lambda_{\rm Com}$  from the particled cell all other cells of the tessellattice are found in a tension state.  
  
 Although whatever changes can occur with a topological ball, the particulate ball's surface area should be invariant. In the tessellattice the surface of cells is in a degenerate state and it is neutral: the number of amplitudes directed inward is practically the same as those directed outward. If spikes are oriented inward we may talk about the negative charge, if spikes are oriented outward -- this could be the positive charge. Thus, fractal transfigurations of the ball's surface, the surface defect $\Delta S$, will consist of exclusively unidirectional surface fractals  $\sigma_n^{(\rm in)}$ or $\sigma_n^{\rm (out)}$. Such a ``chestnut" model of the electric charge in which $N \sim 10^{183}$  [7] spikes protrude inside or outside the particulate ball's surface and if such structure is sustainable, this can be considered as the quantum of the surface fractal. 

Besides, inside this region a polarized coat is also developed, which screens the charge generated by the particle from degenerate cells of the tessellattice. The radius of this polarized coat is equal to Thomason's classical electron radius $r_e = e^2 / (4 \pi \epsilon_0 {\kern 1pt} m c^2)$. Beyond the charged particle's coat, i.e. behind the radius $r_e$ there is no information about its electric properties.

The availability of these two coats are responsible for the manifestation of the fine-structure constant, which appears as a ratio of the coats' radii: $\alpha = r_e/(2 \pi \lambda_{\rm Com})$  [7]. 

Fig. 1$a$ shows how the positively charged particle is screened by polarized cells. In Fig. 1$b$ one can see the charged particle and one of the nearest cells from the polarized coat. 

When such charged particle starts to move, squeezing between the polarized cells, the coat states also migrate. The moving particled cell imposes its coat, i.e. ambient cells adapt to the particled cell in each point of its path.

\begin{figure}
\begin{center}
\includegraphics[width=6in]{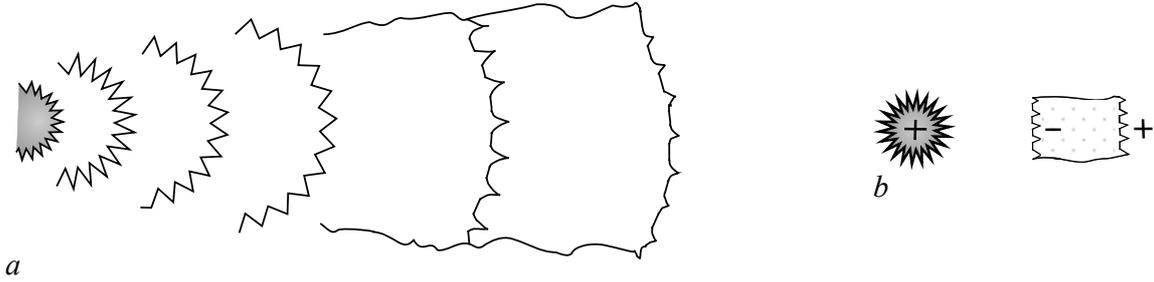}
\caption{ \textit{a} -- the positive charge that polarizes cells in the tessellattice;
\textit{b} -- the charge and the polarized cell.}
\label{Figure 1}
\end{center}
\end{figure}

It is obvious that due to friction, the particle velocity slows down. Indeed, when the particled cell is moving through the tessellattice, it must rub against oncoming cells. In quantum physics we have to call this phenomenon the interaction of the particle with space. Owing to this interaction, the particle emits a cloud of excitations named \textit{inertons}, which transfer the particle's mass and velocity, and the particle gradually loses its speed down to zero. After that, space returns inertons back to the particle and adsorbing them the particle acquires its mass and speed again, and so on. The particle's de Broglie wavelength $\lambda$ plays the role of spatial half a period: in an odd section $\lambda$ the particle emits inertons and finally stops; in the next even section $\lambda$ the space returns these inertons to the particle and the particle, absorbing inertons, restores its velocity. Thus, the particle, at odd sections equal to its de Broglie wavelength $\lambda$, gradually loses its velocity and mass, which are transferred to the emitted inerton cloud. 

However, the presence of the additional electric polarization in the space around the charged particle results in the overlapping of mass and electric properties in emitted inertons. In other words, the particle's inertons become also electrically polarized and hence the moving particle is accompanied by its proper inerton-photon cloud. Below we do not consider the mass component (see Ref. 7 for detail) and will deal only with the electric component of the cloud, i.e. we will consider so-called ``virtual" photons joined to the charged particle.  

At the motion, the surface of both the particled cell and its surrounding cells changes as shown in Fig. 2. Spikes also have to acquire motion -- they can move near their equilibrium positions, such that the radial symmetry changes to the curl symmetry. One can describe a spike by a scalar potential $\varphi$. A declination of the spike, which is not a one-dimensional line but a kind of body, can be depicted by a vector potential \textbf{A}.  So, these two potentials,  $\varphi$ and \textbf{A}, allow us to talk about the appearance of two fields -- electric ${\bf E} = - {\boldsymbol \nabla} \varphi$  and magnetic induction ${\bf B} = {\boldsymbol \nabla} \times {\bf A}$. The origin of the induction {\bf B} is the surface spikes and the particle motion (Fig. 2$b$ and 2$c$).  

Now we can consider the charge $e$ described by the scalar $\varphi_n$ and vector ${\bf A}_n$ fields, where the index $n$ corresponds to the $n$th spike of the particle surface. Dimensionality of these fields corresponds to length. Therefore, their velocities are $\dot \varphi_n $ and $\dot{\bf A}_n$. Let us describe the photon cloud as a single quasi-particle that has the same quantity of spikes as the charge $e$. The photon's nth spike is characterized by scalar $\phi_n$ and vectors ${\boldsymbol \alpha}_n$ fields, respectively.  Now we can construct the Lagrangian density for the motion of the $n$th particle's spike and the nth cloud's spike taking into account their mutual interaction: 
\begin{equation}\label{6}
L_n = C \Big\{ 
\frac12 \dot \varphi_n^2  +    \frac12 \dot {\bf A}_n^2 +  \frac12 \dot\phi_n^2 +  \frac12 {\boldsymbol \alpha}_n^2
+ v \big( \dot{\bf A}_n \cdot {\boldsymbol \nabla} \phi_n   +  \dot {\boldsymbol \alpha}_n \cdot {\bf \nabla} \varphi_n \big) 
 - \frac 12  v^2 ({\boldsymbol \nabla}  \times {\bf A}_n )({\boldsymbol \nabla} \times {\boldsymbol \alpha} )   
 \Big\} ,                            
\end{equation}    
 \noindent  where $C$ is the constant and $v$ is the initial particle velocity. The Lagrangian
density (6) can be split to two independent Lagrangian densities of the particle
and its cloud, respectively [7]:
\begin{equation}\label{7}
L_n^{\rm particle} = C \big\{  \frac12 \dot\varphi_n^2 + \frac 12 \dot {\boldsymbol A}_n^2
+ v  \dot {\bf A}_n \cdot {\boldsymbol \nabla} \varphi_n 
- \frac 12 v^2 ({\boldsymbol \nabla} \times {\bf A})_n^2
\big\},
\end{equation}

\begin{equation}\label{8}
L_n^{\rm cloud} = C\big\{  \frac12 \dot \phi_n^2 + \frac12 \dot { \boldsymbol \alpha}^2_n 
+ v {\kern 1pt} \dot{\boldsymbol \alpha}_n \cdot {\boldsymbol \nabla} \phi_n
- \frac 12  v^2 ({\boldsymbol \nabla} \times {\boldsymbol \alpha}_n^2 )^2
\big\}.
\end{equation}

\noindent
The Euler-Lagrange equations for the Lagrangian densities (7) and (8) result in the wave equations for their variables. Thus, for the particle potentials the equations of motion are

\begin{equation}\label{9}
\ddot\varphi_n - v^2 {\kern 1pt} {\boldsymbol \nabla }^2 \varphi_n = 0,
\end{equation}

\begin{equation}\label{10}
\ddot{\bf A}_n = v^2 {\kern 1pt}{\boldsymbol \nabla}^2 {\bf A}_n = 0.
\end{equation}

\begin{figure}
\begin{center}
\includegraphics[width=11.5in]{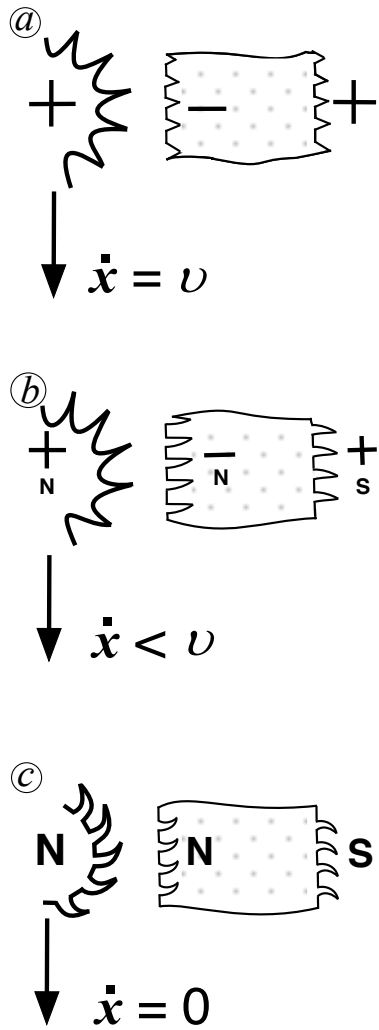}
\caption{The motion of the charge tightly surrounded with polarized cells of the tessellattice. $a$ -- the initial state; $b$ -- intermediate state; $c$ -- the state at the end of the odd section of the particle's de Broglie wavelength, i.e. the particle surface becomes stretched, or tensed and it represents the source of the magnetic induction ${\bf B} = {\boldsymbol \nabla} \times {\bf A}$. In other words, in the final point of de Broglie's wavelength $\lambda$ the particle?s charge e is transformed to the magnetic monopole $g$.}
\label{Figure 2}
\end{center}
\end{figure}

Eq. (9) shows that along the particle path the particle's electric potential obeys the conventional wave equation. Eq. (10) shows that the particle?s vector potential obeys the wave equation as well. 

The consideration above exhibits an interesting aspect: the motions of electric and magnetic properties are carried out in a contra phase, namely, when the electric polarization decreases, the magnetic polarization increases, and vice versa. Although so far, all literature assumes that the electric and magnetic components grow and fall synchronously.  

For a free photon, which is an excitation that migrates hopping from cell to cell (see Fig. 2 to the right) the appropriate Lagrangian looks as follows in the SI units
\begin{equation}\label{11}
L_{\rm photon} =   \frac {\epsilon_0}{2c^2}{\kern 1pt} \dot \varphi^2  + \frac{\epsilon_0}{2} {\kern 1pt}\dot{\bf A}^2
+ \epsilon_0 {\kern 1pt} \dot{\bf A}\cdot {\boldsymbol \nabla} \varphi 
- \frac {\epsilon_0 c^2}{2} ({\boldsymbol \nabla} \times {\bf A})^2.
\end{equation}

\noindent
The equations of motion of the appropriate potentials $\varphi$ and {\bf A} are the same as above (9) and (10), respectively. 

The Lagrangian density of a flux of free photons that interact with a charged particle in standard symbols is 
\begin{equation}\label{12}
L = \frac {\epsilon_0} {2c^2} + \frac{\epsilon_0}{2} {\kern 1pt}\dot{\bf A}^2
+ \epsilon_0 {\kern 1pt} \dot{\bf A}\cdot {\boldsymbol \nabla} \varphi 
- \frac {\epsilon_0 c^2}{2} ({\boldsymbol \nabla} \times {\bf A})^2 
-\rho \cdot (\varphi_0 - \varphi) + \rho {\bf v} \cdot {\bf A}
\end{equation}

\noindent 
where $\rho$  is the charge density, $\varphi_0$ is the reference point of the potential $\phi$ because as in reality we have to consider the difference of the potentials between two points, and {\bf v} is the velocity of the charge. The equations of motions, i.e. the Euler-Lagrange equations for the potentials $\varphi$ and {\bf A}  are as below:

\begin{equation}\label{13}
\frac {1}{c^2} \frac {\partial^2 \varphi}{\partial t^2}  - \nabla^2 \varphi = \frac {\rho}{\epsilon_0},
\end{equation}

\begin{equation}\label{14}
\frac {1}{c^2} \frac {\partial^2 {\bf A}}{\partial t^2}  - \nabla^2 {\bf A} = \mu_0{\kern 1pt} {\bf g}
\end{equation}

\noindent 
where the magnetic monopole ${\bf g} = \rho {\bf v}$ is introduced instead of a conventional ${\bf j} = \rho {\bf  v}$ (see the expression (5)).
       The equations (13) and (14) represent the d'Alambert's form of Maxwell's equations, which include the magnetic monopole as it must be. Namely, the equation (13) shows that the charge density $\rho$ is characterized by the wave movement at which half the spatial period $\lambda$ (the charged particle's de Broglie wavelength) plays the role of a section in which $\rho$ gradually falls down to zero. Since $\rho = e / V$, this means that $e \to 0$. Then in the next (even) section $\lambda$, the value of charge is restored to $e$, and so on. The equation (14) exhibits a similar wave movement for the magnetic monopole $\bf g$. Fig. 2 depicts the process of motion from which it follows that the charge and monopole are moving in their waves in antiphase. Physically this means that the particle's charge $e$ is transformed to the magnetic monopole $g$ in the end of each odd section $\lambda$. Then coming the next even section $\lambda$ the monopole state $g$ turns into the charge state $e$ at the final point of the section, and so on along the whole particle path. 

It is important to emphasize that the electric and magnetic components that are moving have the phase shift $\pi/2$ (Fig. 3), though so far in electrodynamics these two fields have been treated as developed in phase -- they both have been increasing and decreasing, which destroyed the principle of energy conservation in a closed system.

\begin{figure}
\begin{center}
\includegraphics[width=5in]{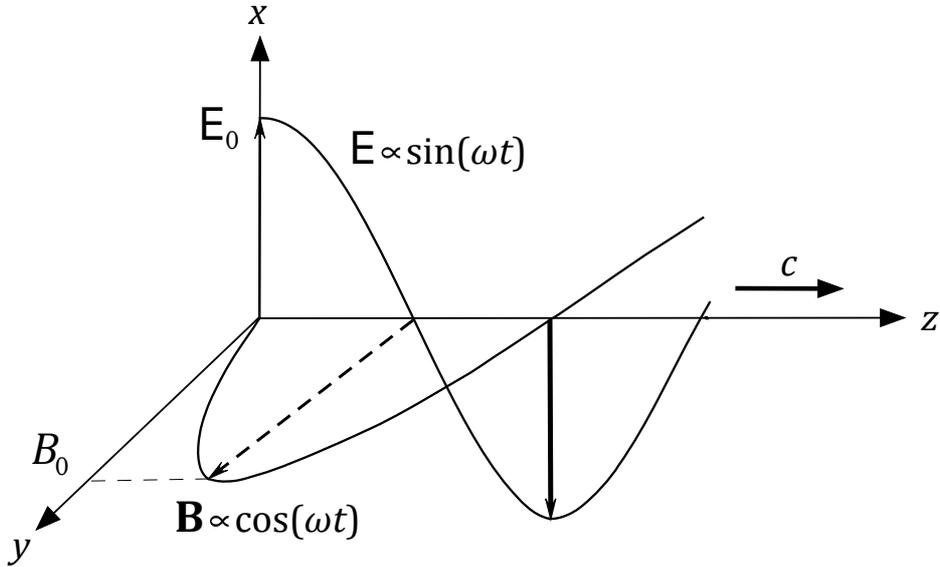}
\caption{Correct presentation of the motion of the electric and magnetic components 
 of the electromagnetic field. }
\label{Figure 2}
\end{center}
\end{figure}

  The proposed Lagrangian density (12) is different from the standard Lagrangian density of the electromagnetic field used in the literature: the expression (12) contains the variable $\varphi$. It has been absent in the conventional Lagrangian density because so far there has not been an understanding of either the nature of the charge or the inner physical processes that lead to Maxwell's equations. In classical electrodynamics the problem has been offset by the supplemented FitzGerald-Lorenz condition 
\begin{equation}\label{15}
{\boldsymbol \nabla} \cdot {\bf A}  + \frac {1}{c^2} \frac {\partial \varphi }{\partial t}  = 0,
\end{equation}

\noindent
which includes the missing derivative $\dot\varphi$.

\section*{4. Conclusion}

The origin of charge, photon, potentials $\varphi$ and {\textbf A}, electric field {\bf E} and magnetic induction {\bf B} were unknown and so far, these important parameters have remained undetermined. The lack of understanding of the submicroscopic aspects of the motion of charge gave rise to a number of different kinds of speculations regarding the presence in nature of a magnetic monopole as a separate fundamental particle or maybe some kind of artificial field formation.

Nevertheless, a detailed consideration based on the theory of real physical space developed by Michel Bounias and the author [8-10, 7] allows us to reveal the structure of space, namely,  the tessellattice of cells with the size estimated as Planck's one $\sqrt{\hbar G/c^3} \sim 10^{-35}$ m where each cell is a primary topological ball of Nature. A particle is created from a cell involving a fractal deformation / inflation of its degenerate volume and also a fractal distortion of its surface. A volumetric fractal deformation is responsible for the appearance of the physical notion of mass and the surface fractal deformation is liable for the physical notion of charge. 

The created charged particle induces its deformation massive and polarization coats around itself, which screen the particle?s characteristics from the degenerated tessellattice. At the motion, the particle rubs against oncoming cells, which leads to the emission of spatial quasiparticles that, being reflected from space, return to the particle. 

The process of emission of inerton-photons becomes periodical and allows the wave behavior of major parameters of the particle -- first of all its mass and charge. The mass $m$ periodically decays to the tension $\Xi$. 
 
 The electric charge e decays to the magnetic monopole g. Besides, components of the charge -- the electric field {\bf E} and magnetic induction {\bf B} -- also periodically change. The particle?s de Broglie wavelength $\lambda$ plays the role of spatial half a period of these oscillations. All the electromagnetic transformations obey Maxwell's equations in which the magnetic monopole takes its place as well:

\begin{eqnarray}
 {\boldsymbol \nabla} \cdot {\bf E}  &=& \frac{\rho}{\epsilon_{0}} , \nonumber  \\
 {\boldsymbol \nabla} \cdot {\bf B} &  = & 0,   \nonumber     \\
 {\boldsymbol \nabla} \times {\bf E} &= &- \frac {\partial {\bf B}} {\partial t},   \nonumber      \\
{ \boldsymbol \nabla} \times {\bf B} & = & \epsilon_0 \mu_0 \frac{\partial {\bf E}} {\partial t} + {\mu_0} {\bf g}. \nonumber 
\end{eqnarray}

\noindent
Mexwetl's equations writlen in the d'Alambart's prestneation, (13) and (14), also naturally include the magnetic monopole.

In this way, we got rid of the mysticism associated with the existence of the magnetic monopole: the monopole appears as a curl state of the moving charge. The conclusion is true for both leptons and quarks. One more inference [7]: the neutrino is the magnetic monopole of the positively charged lepton, i.e. positron, which has been emitted at a non-adiabatic reaction of transformation (note in the monopole state the particle possesses neither electric charge, nor mass). The neutrino is the classical example of a stable magnetic monopole $g$: it possesses neither initial mass, nor charge; the neutrino carries only the magnetic monopole  $g$ and tension $\Xi$ of the initial positron. Nevertheless, the neutrino has a small inertial mass associated with the process of motion through the tessellattice as has been described in book [7].

\bigskip
\textbf{References}

\begin{enumerate}
	\item G. Lochak, The equation of a light leptonic magnetic monopole and its
experimental aspects, \textit{Z. Naturfhrssh}.\textbf{ A62}, No. 5-6, 231-246
(2007); also arXiv:0801.2752 [quant-ph].
	\item P. A. M.  Dirac, Quantized singularities in the electromagnetic field,
\textit{Proc. Roy. Soc. London, Ser A }\textbf{133}, No. 821,\textbf{ }60-72
(1931).
	\item J. Preskil, Magnetic monopoles, \textit{Ann. Rev. Nucl. Part. Sci}. \textbf{34},
No. 1, 461-530 (1984).
	\item Y. M. Shnir, Magnetrc monopoles, Springer-Verlag, Berlin, Heidelberg (2005).
	\item S. Khademi, M. Shahsavari and A. M. Saeed, Non-singular magnetic monopoles,
arXiv:0608051 [physics].
	\item G. A. Gonalo and R. E. Zich, Magnetic monopoles and Maxwell's equations in N
dimensions, \textit{2013 International Conference on Electromagnetics in Advanced
Applications (ICEAA)}, \textit{IEEE Xplore} (2013),
\href{https://ieeexplore.ieee.org/document/6632510}{\uline{https://ieeexplore.ieee.org/document/6632510}}.
	\item V. Krasnoholovets, \textit{Structure of Space and the Submircroscopic
Deterministic Concept of Physics}, Apple Academic Press, Oakville, Canada;
Waretuwn, USA (2017).
	\item M. Bounias and V. Krasnoholovets, Scanning the structure of ill-known spaces:
Part 1. Founding principles about mathematical constitution of space.
\textit{Kibernetes: The International Journal of Systems \& Cybernetics}
\textbf{32}, No. 7/8, pp. 945-75 (2003) (a special issue on new theories about time and space, Eds.: L. Feng, B. P. Gibson and Yi Lin); also  arXiv.org/physins/0211096.
	\item M. Bounias and V. Krasnoholovets, Scanning the structure of ill-known spaces:
Part 2. Principles of construction of physical space, \textit{ibid.} \textbf{32},
No. 7/8, 976-1004 (2003); also arXiv.org/physics/0212004.
	\item M. Bounias and V. Krasnoholovets, Scanning the structure of ill-known spaces:
Part 3. Distribution of topological structures at elementary and cosmic scales,
\textit{ibie}. \textbf{32}, No. 7/8, 1005-1020 (2003); arXiv.org/physccs/0301049.
\end{enumerate}

\end{document}